\documentclass[prx,twocolumn,superscriptaddress,longbibliography,aps]{revtex4-1}

\usepackage[T1]{fontenc}
\usepackage[utf8]{inputenc}
\usepackage{lmodern}
\usepackage[sumlimits]{amsmath} 
\usepackage{amssymb}
\usepackage{yhmath}
\usepackage{graphicx}
\usepackage{epstopdf}
\usepackage{latexsym}
\usepackage{color}
\usepackage{nicefrac}
\usepackage{dsfont} 
\usepackage{wasysym} 
\usepackage{stmaryrd}
\usepackage{hyperref} 
\usepackage{xcolor}
\usepackage{multirow}
\usepackage{soul}
\usepackage{adjustbox}
\usepackage{cancel}

\usepackage{esint}
\usepackage{tikz}

\definecolor{airforceblue}{rgb}{0.36, 0.54, 0.66}

\newcommand{\be}{\begin{equation}}
\newcommand{\ee}{\end{equation}}
\newcommand{\bea}{\begin{eqnarray}}
\newcommand{\eea}{\end{eqnarray}}

\usepackage{verbatim}
\usepackage{stmaryrd}
\usepackage[sumlimits]{amsmath} 
\usepackage{dsfont} 
\usepackage{caption}
\usepackage{cancel}
\usepackage{url}

\usepackage{subcaption}

\usepackage{sansmath}
\DeclareFontEncoding{LGR}{}{}
\DeclareSymbolFont{sfgreek}{LGR}{cmss}{m}{n}
\SetSymbolFont{sfgreek}{bold}{LGR}{cmss}{bx}{n}
\DeclareMathSymbol{\sxi}{\mathord}{sfgreek}{`x}

\newcommand{\mb}{\mathbf}
\newcommand{\bs}{\boldsymbol}
\newcommand{\mf}{\mathfrak}

\newcommand{\mt}{\mathtt}
\renewcommand{\ul}{\underline}

\usepackage{scalerel}

\usepackage{accsupp}

\begin{document}

\title{Phonon Thermal Hall Conductivity from Scattering with Collective Fluctuations}

\author{L\'eo Mangeolle}
\affiliation{\'{E}cole Normale Sup\'{e}rieure de
  Lyon, CNRS, Laboratoire de physique, 46, all\'{e}e
d'Italie, 69007 Lyon}
\author{Leon Balents}
\affiliation{Kavli Institute for Theoretical Physics, University of
  California, Santa Barbara, CA 93106-4030}
\affiliation{Canadian Institute for Advanced Research, Toronto, Ontario, Canada}
\author{Lucile Savary}
\affiliation{\'{E}cole Normale Sup\'{e}rieure de
  Lyon, CNRS, Laboratoire de physique, 46, all\'{e}e
d'Italie, 69007 Lyon}
\affiliation{Kavli Institute for Theoretical Physics, University of
California, Santa Barbara, CA 93106-4030}

\date{\today}
\begin{abstract}
  Because electrons and ions form a coupled system, it is a priori
  clear that the dynamics of the lattice should reflect symmetry
  breaking within the electronic degrees of freedom.  This has been
  recently clearly evidenced for the case of time-reversal and mirror symmetry
  breakings by observations of a large phononic thermal Hall effect in
  many strongly correlated electronic materials. The mechanism by which
  time-reversal breaking and chirality is communicated to the lattice
  is, however, far from evident.  In this paper we discuss how this
  occurs via many-body scattering of phonons by collective modes: a
  consequence of non-Gaussian correlations of the latter modes.  We
  derive fundamental new results for such skew (i.e.\ chiral)
  scattering and the consequent thermal Hall conductivity. We
    emphasize that these
    results apply to any collective variables in any phase of matter,
    electronic, magnetic or neither, highly fluctuating and
    correlated, or not. As a proof of principle, we compute
  general formulae for the above quantities for ordered
  antiferromagnets.  From the latter we obtain the scaling behavior of
  the phonon thermal Hall effect in clean antiferromagnets.  The calculations
  show several different regimes and give quantitative estimates of similar
  order to that seen in recent experiments.
\end{abstract}

\maketitle

\section{Introduction}
\label{sec:introduction}

Thermal conductivity is the most ubiquitous transport coefficient,
being well-defined in any system with sufficiently local interactions,
irrespective of the nature of the specific low energy degrees of
freedom.  It is particularly important therefore in systems for which
charge transport is either strongly suppressed (i.e.\ insulators) or
singular (i.e.\ superconductors).   Moreover, the thermal {\em Hall}
conductivity plays a particularly special role in the theory of exotic
topological phases, as it can be related to the chiral central charge,
to the presence of edge modes, etc.  For all of the above reasons,
experiments on thermal conductivity have played a pre-eminent role in
establishing the nature of the most interesting strongly correlated
states of matter.   A few notable examples are the observation of
metallic-like transport in an organic spin
liquid \cite{dmit131sciencemag}, a quantized thermal Hall effect in the
Kitaev material $\alpha$-RuCl$_3$ \cite{kasahara2018majorana}, taken as
evidence for Majorana fermion edge states, and an exceptionally large
and yet unexplained thermal Hall effect in under-doped cuprate
high-temperature superconducting materials \cite{grissonnanche2019giant,boulanger2020thermal}.

\begin{figure}[htbp!]
  \centering
  \includegraphics[width=\columnwidth]{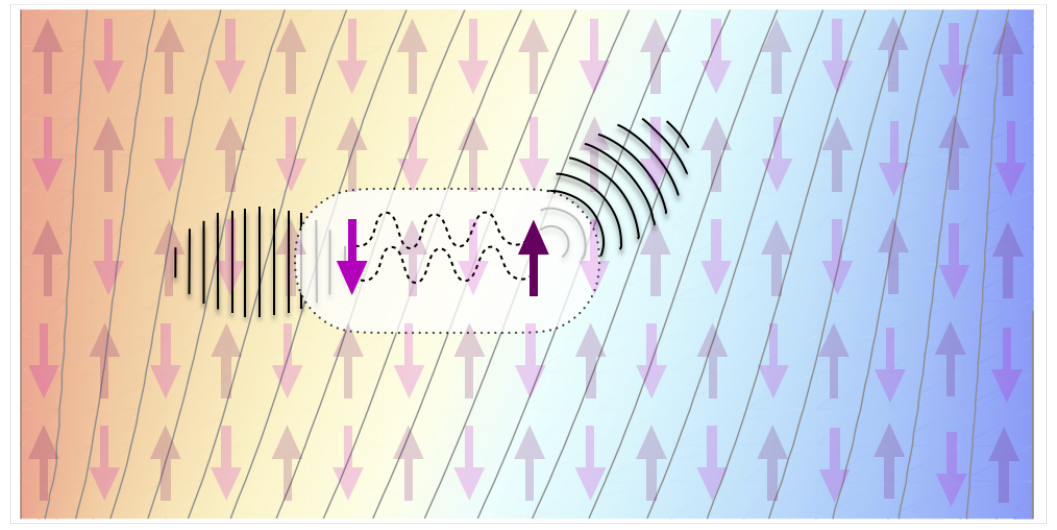}
  \caption{Illustration of a scattering mechanism responsible for a
    Hall effect. Only scattering processes which involve at least two (virtual)
  collisions with collective fluctuations can contribute to a Hall
  effect.}
  \label{fig:scattering}
\end{figure}

Arguably the Achilles heel of thermal conductivity measurements is the
contribution of lattice vibrations/phonons to heat transport.  Phonons
are present in any solid, and indeed, except at very low temperature,
usually dominate the thermal properties of materials.  A common
approach to this fact is to attempt to separate electronic and lattice
contributions by some subtraction scheme, for example based on
measuring two electronically different but vibrationally similar
analog materials, or on dependences on temperature, field, etc., which
might be attributed uniquely to only one of the lattice or electronic
degrees of freedom.  Of particular significance in this regard is the
thermal Hall effect, which by Onsager relations can only exist when
time-reversal symmetry is broken \cite{casimir}.
The Hall conductivity is captured by
the antisymmetric components of the thermal conductivity tensor
$\boldsymbol{\kappa}$, namely
\begin{equation}
  \label{eq:28}
  \kappa_H^{\mu\nu}(T,\bs{H},\cdots)=(\kappa^{\mu\nu}-\kappa^{\nu\mu})/2.
\end{equation}
It has been often assumed that the charge neutrality of phonons and the large ionic mass are sufficient to
prevent them from coupling effectively to internal or external magnetic
fields, and therefore that large thermal Hall signals must arise
uniquely from the electrons in a material. Many recent theoretical
works have thus focused on the thermal
  Hall conductivity of spin excitations \cite{katsura2010theory, doi:10.7566/JPSJ.86.011007,
    han2019consideration, samajdar2019thermal, PhysRevResearch.2.033283},
  in particular spin waves \cite{matsumoto2011rotational, murakami2017thermal, PhysRevB.99.014427, PhysRevB.104.075121}.

Recent experiments, however, have conclusively shown that this assumption is incorrect,
via the simplest and most persuasive of arguments \cite{onose2019,behnia2020,taillefer-simple}.  In particular,
studies of materials which are electronically (or magnetically)
two-dimensional have observed that the thermal Hall conductivity is
three-dimensional, and remains large when the thermal
current within the sample is normal to the two-dimensional planes
\cite{grissonnanche2020chiral}. One has no choice but to conclude that the transported heat is carried by phonons.

The problem posed by these observations is then to understand how
lattice vibrations ``sense'' time reversal symmetry breaking.  This
must indeed be by an indirect process, as ultimately it is the
electrons which interact directly and significantly with magnetic
fields.  In principle there are two broad ways in which the transfer
of information, i.e. the breaking of time-reversal symmetry, can occur from electrons to the lattice.
First, it can occur via the quasi-adiabatic adaptation of electronic states to slow
phonon motions, which may modify the phonon
dispersion relations and generate dynamical Berry phases 
\cite{PhysRevLett.96.155901, PhysRevLett.100.145902,
  PhysRevLett.123.167202,PhysRevLett.124.167601}. While this is certainly possible in
principle, numerous estimates indicate that this mechanism is unlikely
to explain the large magnitude of thermal Hall signals seen in
experiments.  The second type of information transfer, depicted in Fig.~\ref{fig:scattering}, is through
scattering of phonons from the electronic modes, which can be
``chiral'' when the latter break time-reversal and reflection
symmetries.  In the electrical anomalous Hall effect, such ``skew
scattering'' is known to dominate in the most highly conducting
samples \cite{saito2019berry}, and for similar reasons, we expect it to
do so for heat transport when thermal conductivity is large.

\begin{figure}[htbp]
  \centering
  \begin{subfigure}[b]{0.825\columnwidth}
    \includegraphics[width=\columnwidth]{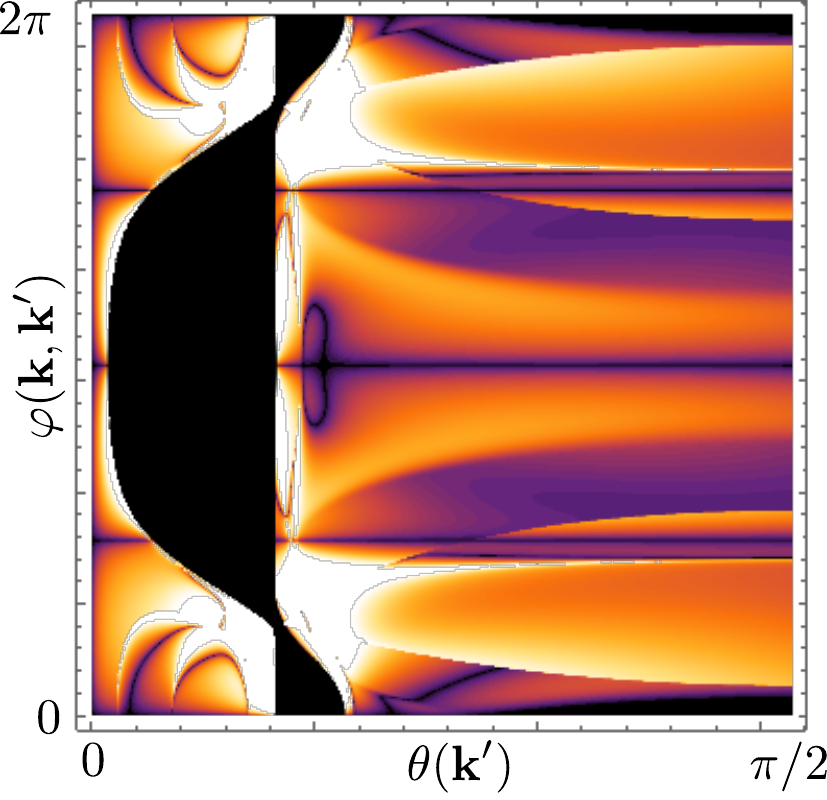}
  \end{subfigure}
  \begin{subfigure}[b]{0.16\columnwidth}
    \includegraphics[width=\columnwidth]{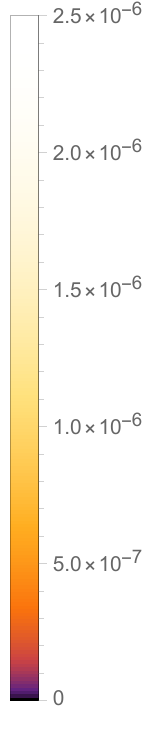}
    \end{subfigure}
    \caption{Calculated skew-scattering rate
      $\mf W^{\ominus,-+}_{n\mb k n'\mb k'}/\gamma_0$ (see
      Eqs.~(\ref{eq:8},~\ref{eq:17})) to transfer a phonon in mode
      $n'\mb{k}'$ into mode $n\mb{k}$ induced by coupling the lattice
      to a two dimensional antiferromagnet.  The density plot shows
      the angular dependence as a function of
      $\theta(\mb k')\in [0,\pi/2]$ (horizontal axis) and
      $\varphi(\mb k,\mb k')=\phi(\mb k')-\phi(\mb k)$ (vertical axis)
      for fixed $|\mb k'|=\mt{0.8}/\mf a$, $k_x=\mt{0.2}/\mf a$,
      $k_y = 0$, $k_z = \mt{0.1}/\mf a$,
      $\bs m_0=\mt{0.05}\hat{\bs z}$ and temperature
      $T=\mt{0.5}T_0$. Here $\mathfrak{a}$ is the in-layer lattice
      spacing, $\phi(\mathbf{k}^{(')})$ and $\theta(\mathbf{k}^{(')})$
      are the azimuthal and polar angles of
      $\mathbf{k}^{(')}$, defined in the usual way. Note that the colorbar is not
      scaled linearly.}
  \label{fig:WH}
\end{figure}

With this in mind, it is critical to ask how time-reversal symmetry
breaking of electronic degrees of freedom is communicated {\em via
  scattering} to phonons in clean systems.  We assume perturbative
coupling of some set of collective fields $Q$ to the lattice, which is
generally valid away from the limit of polaron formation
\cite{holstein1959-I, holstein1959-II}.  To account for the diversity of different electronic and magnetic phases being
studied, we allow the fields $Q$ to be {\em general}, restricted only by
the requirements of unitarity of quantum mechanics and equilibrium.
We show that the full scattering data needed to understand the thermal
conductivity (both longitudinal and Hall components) can be obtained
from the time and space-dependent correlation functions of the $Q$
fields.  Crucially, we show that {\em the standard two-point
  correlation functions of $Q$ give vanishing contributions to skew
  scattering and the Hall effect}.  Consequently the skew scattering
can be attributed entirely to {\em non-Gaussian} fluctuations of the
collective modes. 
This is a challenge theoretically (because as we
discuss below, beyond-Gaussian fluctuations are significantly more
complex than Gaussian ones) but also an opportunity.  The absence of
lower-order contributions to skew scattering means that the latter
provides a direct probe of non-Gaussianity, which does not require any
subtraction!  This suggests the prospect of using measures of skew
scattering of phonons, such as the thermal Hall effect, as a means to
interrogate the rich higher order correlations of electronic modes.

In this paper, we identify the corresponding higher order
correlation functions which relate the multi-phonon scattering rates
to the fluctuations of the collective modes.  These are complicated
objects which depend upon several time and space coordinates, or
equivalently multiple frequencies and wavevectors.  We show how to
extract the essentially antisymmetric part of these correlations which
uniquely contribute to the Hall effect, using symmetry and detailed
balance relations, which generalize well-known and ubiquitously
important laws that are used to analyze two-point correlations
throughout physics \cite{onsager, squires_neutron_scatt, buettiker}. 
This provides a recipe which can be applied in
diverse systems, telling what must be known about the electronic modes
coupled to the lattice and how to use that data to obtain an
understanding of the thermal Hall effect of phonons.  Notably, the
results are valid irrespective of the nature of the phase of matter
hosting the collective fields: it may be strongly fluctuating, highly
correlated, or even have no quasiparticles at all.  This contrasts
greatly with prior theories of phonon skew scattering which are based on
very specific models of electronic modes \cite{Lovesey_1972,
  PhysRevB.8.2130}.

To demonstrate the methodology and as a proof of principle, we also
apply the general results to the case of an ordered antiferromagnet,
in which case the $Q$ fields correspond to magnetic fluctuations which
can be decomposed into composites of magnons.  The result is a richly
structured skew scattering rate, visualized in Fig.~\ref{fig:WH}.
Validating the general formulation, we obtain a non-vanishing thermal
Hall effect when all the symmetry criteria (which we establish) are
satisfied, and we explicitly show that within a minimal model of an
antiferromagnet with strictly two-dimensional magnetic correlations,
the thermal Hall effect is three-dimensional and its magnitude is
roughly independent of whether the thermal currents are within or
normal to the magnetic planes.

\section{Scattering and correlation functions}
\label{sec:scatt-corr-funct}

In this section, we present the main results for the scattering rates
of phonons due to collective modes.  We limit the discussion here to the
simplest case in which the coupling is linear in phonon
creation/annihilation operators $a_{n\mb{k}}^\dagger,
a_{n\mb{k}}^{\vphantom\dagger}$.   Then, the coupling Hamiltonian is
\begin{equation}
\label{eq:1}
H' = \sum_{n\mb{k}}  \left( a_{n\mb{k}}^\dagger
  Q_{n\mb{k}}^\dagger+a_{n\mb{k}}^{\vphantom\dagger} Q_{n\mb{k}}^{\vphantom\dagger}\right),
\end{equation}
where $Q_{n\mb{k}}$ describes the collective mode arising from
electronic degrees of freedom, coupled to the $n^{\rm th}$ phonon
polarization.  For brevity, we subsume any electron-phonon coupling
constant into $Q_{n\mb{k}}$.  We carry out a perturbative analysis of
$H'$, so $Q_{n\mb{k}}$ may be regarded as ``small''.

\subsection{Formulation}
\label{sec:formulation}

Our aim is to calculate the necessary terms in the collision integral $\mathcal{C}_{n\mb{k}}$
of the  phonon Boltzmann equation,
\begin{equation}
  \label{eq:2} \partial_t \overline{N}_{n\mathbf{k}}+\bs{v}_{n\mathbf{k}}\cdot{\boldsymbol{\nabla}}_\mathbf{r}\overline{N}_{n\mathbf{k}}
= \mathcal C_{n\mathbf{k}}[\{\overline{N}_{n'\mathbf{k}'}\}]
\end{equation}
where $\overline{N}_{n\mb{k}}$ is the non-equilibrium average
occupation number of phonons in polarization mode $n$ and
quasi-momentum $\mb{k}$ with velocity 
$\bs{v}_{n\mathbf{k}}=\boldsymbol{\nabla}_\mathbf{k}\omega_{n\mathbf{k}}$,
where $\omega_{n\mathbf{k}}$ is the $(n,\mathbf{k})$ phonon dispersion relation.  Once the collision integral is known, the
Boltzmann equation can be solved in a standard manner by linearizing
around the equilibrium distribution, to obtain the non-equilibrium
change and thereby the transport current to linear order in the
temperature gradient.  

We now summarize the method used to obtain the collision integral from
the microscopic quantum dynamics and Eq.~\eqref{eq:1}.  The basic
procedure is to determine the {\em many body} transition rate between
microstates in the combined phonon-electron system using the
scattering matrix ($T$) expansion, and from there, use the equilibrium
distribution for the electronic subsystem to evaluate the rate of
change of the mean occupation probabilities of phonon states that
enter the Boltzmann equation.

We begin with the Born expansion \cite{landau2013quantum}:
\begin{equation}
\label{eq:3}
  T_{\mathtt {i\rightarrow f}}=T_{\mathtt{fi}}=\langle \mathtt{f}|H'|\mathtt{i}\rangle 
  +\sum_\mathtt{n}\frac{\langle\mathtt{f}|H'|\mathtt{n}\rangle\langle\mathtt{n}|H'|\mathtt{i}\rangle}{E_\mathtt{i}-E_\mathtt{n}+i\eta}+\cdots,
\end{equation}
where the $|\mathtt{i}\rangle, |\mathtt{f}\rangle, |\mathtt{n}\rangle$ states are
product states in the $Q$ (index $s$) and phonon (index $p$) Hilbert space,
$|\mathtt{g}\rangle=|g_s\rangle|g_p\rangle$ for
$g=i,f,n$, and $E_{\mathtt{g}}$ is the energy of the unperturbed
Hamiltonians of the $Q$ and phonons in state ${\mathtt{g}}$. $\eta\rightarrow0^+$
is a small regularization parameter. 

The rate of transitions
from state $\mathtt{i}$ to state $\mathtt{f}$ is obtained using Fermi's golden rule,
\begin{equation}
\label{eq:4}
  \Gamma_{\mathtt {i\rightarrow f}} =\frac{2\pi }\hbar
\,|T_{\mathtt{i\rightarrow f}}|^2\delta(E_{\mathtt{ i}}- E_{\mathtt  {f}}).
\end{equation}
Note that $\Gamma_{\mathtt {i\rightarrow f}}$ is a transition rate in
the full combined phonon-$Q$ system.  By assuming equilibrium for the
electronic modes, we obtain the transition rates within the phonon subsystem,
\begin{equation}
\label{eq:5}
  \tilde\Gamma_{i_p\rightarrow f_p}=\sum_{i_sf_s}\!\Gamma_{\mathtt{i}\rightarrow\mathtt{f}}\,p_{i_s},
\end{equation}
with $p_{i_s}=\frac 1 {Z_s} e^{-\beta E_{i_s}}$.
This in turn determines the
collision integral through the master equation
\begin{equation}
\label{eq:6}
  \mathcal{C}_{n\mathbf{k}}=\sum_{i_p,f_p}\tilde\Gamma_{i_p\rightarrow f_p}\left(N_{n\mathbf{k}}(f_p)-N_{n\mathbf{k}}(i_p)\right)p_{i_p},
\end{equation}
where $p_{i_p}=\sum_{i_s}p_{\mathtt{i}}$, where $p_{\mathtt{i}}$ is
the probability to find the system in state $\mathtt{i}$.

\subsection{Result}
\label{sec:result}

To carry out the above procedure, we first express the microscopic
processes generated in the Born expansion, Eq.~\eqref{eq:3}, and
insert them into the square in Eq.~\eqref{eq:4}.  Then the sums over
electronic states $i_s,f_s$ in Eq.~\eqref{eq:5} can be converted into
dynamical multi-time correlation functions of the $Q$ operators.  The
corresponding technical manipulations are described in Appendix~\ref{sec:from-inter-terms}.
The leading result for the longitudinal conductivity (symmetric part
of the tensor, $\kappa_L^{\mu\nu}=(\kappa^{\mu\nu}+\kappa^{\nu\mu})/2$) is dominated by
diagonal scattering (absorption or emission of a single phonon), and
is given by
\begin{equation}
  \label{eq:25}
  \kappa_L^{\mu\nu}=\frac{\hbar^2}{k_BT^2}\frac{1}{V}\sum_{n\mathbf{k}}\frac{\omega_{n\mathbf{k}}^2
    v^\mu_{n\mathbf{k}}v^\nu_{n\mathbf{k}}}{4D_{n\mathbf{k}}\sinh^2(\beta\hbar\omega_{n\mathbf{k}}/2)},
\end{equation}
where $V$ is the volume of the system, $\mu,\nu=x,y,z$ and $D_{n\mathbf{k}}$ is the longitudinal scattering rate (see Eq.~\eqref{eq:11}).
For the Hall effect, the important contributions are those from the
second order terms in the Born expansion, which generate processes in
which a phonon is scattered from one state to another, or in which
pairs of phonons are created or annihilated.  Using Eq.~\eqref{eq:6}
then leads to off-diagonal terms in the collision integral,
i.e.\ contributions to $\mathcal{C}_{n\mb{k}}$ proportional to
$N_{n'\mb{k}'}$ with $n'\mb{k}' \neq n\mb{k}$.  The desired ``skew''
scattering contributions, roughly speaking, correspond to processes in
which $\mb{k}$ is preferentially deflected ``to the right'' of $\mb{k}'$, for
example.
\begin{widetext}
More precisely, we define anti-symmetric scattering rates $\mathfrak W^{\ominus,+,q}_{n\mathbf k,n'\mathbf
   k'}$ ($q=\pm$) in such a
way that they control the anti-symmetric (Hall) part of the thermal
conductivity tensor, $\kappa^{\mu\nu}_H = -\kappa^{\nu\mu}_H$:
 \begin{equation}
   \label{eq:7}
    \kappa_H^{\mu\nu} =  \frac{\hbar^2}{k_B T^2} \frac{1}{V} 
    \sum_{n\mathbf{k}n'\mathbf{k}'}
    J^\mu_{n\mathbf{k}}\frac{e^{\beta\hbar\omega_{n\mathbf{k}}/2} }{2D_{n\mathbf{k}}}
    \Bigg (\frac 1 {N_{\rm uc}}\sum_{q=\pm}\frac{\left(e^{\beta\hbar\omega_{n\mathbf{k}}}-e^{q\beta\hbar\omega_{n'\mathbf{k}'} }\right)\,\mathfrak W^{\ominus,+,q}_{n\mathbf k,n'\mathbf
   k'}}{\sinh(\beta\hbar\omega_{n\mathbf  k}/2)\sinh(\beta\hbar\omega_{n'\mathbf k'}/2)}
\Bigg ) \frac{e^{\beta\hbar\omega_{n'\mathbf{k}'}/2} }{2 D_{n'\mathbf{k}'}}J^\nu_{n'\mathbf{k}'} ,
\end{equation}
where, for $q,q'=\pm1$,
\begin{equation}
  \label{eq:8}
  \mathfrak{W}^{\ominus,qq'}_{n\mathbf{k}n'\mathbf{k}'}
  =\frac{2 N_{\rm uc}}{\hbar^4}\mathfrak{Re}\int_{t,t_1,t_2}\! e^{i[\Sigma_{n\mb k n'\mb k'}^{q, q'}
  t+\Delta_{n\mb k n'\mb k'}^{q, q'}(t_1+t_2)]}
  \textrm{sign}(t_2) \left\langle\left[
    Q^{-q}_{n\mathbf{k}}(-t-t_2),Q^{-q'}_{n'\mathbf{k}'}(-t+t_2)\right]\left\{
    Q^{q'}_{n'\mathbf{k}'}(-t_1),Q^{q}_{n\mathbf{k}}(t_1)\right\}\right\rangle,
\end{equation}
and
\begin{equation}
  \label{eq:11}
    D_{n\mathbf k} = -\frac {1}{\hbar^2} \int dt\, e^{-i\omega_{n\mathbf{k}}t}\left \langle [Q_{n\mathbf k}^{\vphantom\dagger} (t),
    Q^\dagger_{n\mathbf{k}}(0)]\right\rangle_\beta + \breve D_{n\mb{k}}.
\end{equation}
Here $\langle\cdots\rangle_\beta$
denotes the expectation value at inverse temperature $\beta$, $\mu,\nu=x,y,z$, and we defined the phonon current
    $J^\mu_{n\mathbf{k}}=N_{n\mathbf{k}}^{\rm
      eq}\,\omega_{n\mathbf{k}}v^\mu_{n\mathbf{k}}$,
   and $N_{n\mathbf{k}}^{\rm eq}$ is the number of phonons in mode
   $n\mathbf{k}$ in thermal equilibrium.  Eq.~\eqref{eq:11} gives the
   leading order result for the diagonal scattering rate $D_{n\mb{k}}$, which enters
   Eq.~\eqref{eq:7}.   In general it includes contributions $\breve
   D_{n\mb{k}}$ from other scattering channels (e.g.\ impurities) and
   higher-order contributions.  In Eq.~\eqref{eq:8}  we introduced the
   notation $Q^+_{n\mb{k}} = Q^\dagger_{n\mb{k}}$ and $Q^-_{n\mb{k}} =
   Q_{n\mb{k}}$, as well as $\Sigma_{n\mb k n'\mb k'}^{q,
   q'}=q\omega_{n\mathbf{k}}+q'\omega_{n'\mathbf{k}'}$ and 
$\Delta_{n\mb k n'\mb k'}^{q,
  q'}=q\omega_{n\mathbf{k}}-q'\omega_{n'\mathbf{k}'}$. Here and in the following, lower case latin $q$ (with or without primes or subscripts) is used to indicate a particle/hole index taking values $\pm 1 = \pm$.
\end{widetext}
Eqs.~(\ref{eq:7},\ref{eq:8}) constitute the central result of this
paper.  They give a general formula for the skew scattering rate and
the thermal Hall conductivity, given Eq.~\eqref{eq:1}, assuming small
Hall angle (a condition which is nearly always true), valid in any
dimension.  Even more general formulae valid when electronic modes are
coupled to both linear and quadratic functions of the phonons will be
given in a separate publication \cite{phononcompanion}.  These results can be applied to any
material provided the non-Gaussian correlations of the collective
degrees of freedom corresponding to $Q_{n\mb{k}}$ are known.
   
Considerable structure is encoded in Eq.~\eqref{eq:8}.  It is
straightforward to show that the skew scattering vanishes if
$Q_{n\mathbf{k}}$ is taken to be Gaussian: in this case,  Wick's theorem
is obeyed, and its application to Eq.~\eqref{eq:8} implies
that $\mf W^{\ominus}_{n\mb{k}n'\mb{k}'}$ is zero if $(n,\mb{k})\neq
(n',\mb{k}')$.  Hence, non-trivial contributions to the skew scattering
arise entirely from non-Gaussian correlations. 
Physically, $\mf W^{\ominus,++}$ (resp.\ $\mf W^{\ominus,--}$)
corresponds to scattering processes where two phonons are emitted
(resp.\ absorbed), and $\mf W^{\ominus,+-},\mf W^{\ominus,-+}$ to
processes where one phonon is emitted and one is absorbed.  The
contribution to the Hall conductivity has been carefully isolated so
that the rate obeys the ``anti-detailed balance'' relation:
\begin{equation}
\label{eq:9}
  \mathfrak W_{n\mathbf k n'\mathbf k'}^{\ominus,qq'} = - ~
  e^{-\beta(q\omega_{n\mathbf k}+q'\omega_{n'\mathbf k'})}~\mathfrak
  W_{n\mathbf k n'\mathbf k'}^{\ominus,-q-q'},
\end{equation}
as well as
\begin{equation}
  \label{eq:10}
  \mathfrak{W}^{\ominus,qq'}_{n\mathbf{k}n'\mathbf{k}'}=\mathfrak{W}^{\ominus,q'q}_{n'\mathbf{k}'n\mathbf{k}}.
\end{equation}
The combination of the commutator and anti-commutator in
Eq.~\eqref{eq:8} ensures the validity of these relations.

\section{Application to an ordered antiferromagnet}
\label{sec:appl-an-order}

We now provide an application of the above results to the specific
case of an insulating antiferromagnet.  This is important as a proof
of principle to confirm that the general formula in Eq.~\eqref{eq:8}
indeed results in a non-vanishing Hall effect of phonons from skew
scattering.  It is also a relevant test case as it corresponds to the
situation in many recent experiments, and is perhaps the simplest
situation in which time-reversal symmetry breaking of spins is
communicated to phonons in an insulator.

To model the antiferromagnet, we employ a spin wave description, and
for concreteness assume the spin correlations are purely
two-dimensional: each layer of spins is presumed completely
independent.  The latter assumption is not essential but it is
illustrative: using it we demonstrate that even when spin correlations
are confined to two dimensions, their influence can lead to thermal
Hall conductivity with heat current oriented perpendicular to those
layers.  In any case, the general formulae in the first subsection below can be
easily modified for the case of three-dimensional spin waves.

\subsection{Formulation and general results within linear spin wave
  theory}
\label{sec:form-gener-results}

 The spin waves are described by magnon operators
$b_{\ell\mb{k},z}^\dagger$ ($b_{\ell\mb{k},z}^{\vphantom\dagger}$), which
create (annihilate) a magnon in branch $\ell$ with momentum $\mb{k}$
in layer $z\in\mathbb{N}$, whose Hamiltonian is
\begin{equation}
  \label{eq:12}
  H_m = \sum_{\ell,\mb{k},z} \Omega_{\mb{k},\ell}
  b_{\ell\mb{k},z}^\dagger b_{\ell\mb{k},z}^{\vphantom\dagger}.
\end{equation}
Note that the effect of a magnetic field is already included in $H_m$,
i.e.\ here the spin wave modes are based on an expansion around the
spin order {\em including} the effect of the field.   The collective
modes $Q^q_{n\mb{k}}$ can be expanded in a series in the spin wave
operators, and the dominant contribution to scattering comes from
second order \footnote{A linear term in magnon operators is generically
  present but does not contribute significantly to scattering due to
  phase space constraints.}:
\begin{equation}
\label{eq:13}
 Q^q_{n\mathbf k}  =  \frac{1}{\sqrt{N_{\rm uc}}}\sum_{\substack{\mathbf p,\ell_1,\ell_2\\q_1,q_2,z} } 
  \!\!\!\mathcal B^{n,\ell_1,\ell_2|q_1 q_2 q}_{\mb k;\mb p} ~ e^{ik_z z} b^{q_1}_{\ell_1,\mathbf
                                                          p+\frac{q}{2}\mb
                                                          k,z}
                                                          b^{q_2}_{\ell_2,-\mb
                                                          p
                                                          +\frac{q}{2}\mb k,z} ,
\end{equation}
where $q=\pm1$ and the sums run over $\mathbf{p}$ in the 2d
  Brillouin zone, $q_{1,2}=\pm1$, $z\in\mathbb{N}$ and $\ell_{1,2}$
  over the magnon branches. Here we defined the notations
\begin{align}
\label{eq:14}
  b_{\ell,\mb p,z}^+ = b^\dagger_{\ell,\mb p,z}, &&  b_{\ell,\mb p,z}^- = b^{\vphantom\dagger}_{\ell,-\mb p,z}.
\end{align}
Note the minus sign in the momentum in the second relation.  This
means generally that
$\left(b^q_{\ell,\mb p,z}\right)^\dagger = b^{-q}_{\ell,-\mb{p},z}$.
To make the coefficients unambiguous, we choose the symmetrized form
$\mathcal B^{n,\ell_1,\ell_2|q_1 q_2 q}_{\mb k;\mb p} = \mathcal
  B^{n,\ell_2,\ell_1|q_2 q_1 q}_{\mb k;-\mb p}$.  
Demanding that $Q^+_{n\mb k} = (Q_{n\mb k}^-)^\dagger$ implies that
$ \mathcal B^{n,\ell_1,\ell_2|q_1 q_2 +}_{\mb k;\mb p} = \left(
  \mathcal B^{n,\ell_2,\ell_1|-q_2 -q_1 -}_{\mb k;\mb p}\right)^*$.  

Eq.~\eqref{eq:12} contains the
energy dispersion $\Omega_{\mb{k},\ell}$ of the spin waves but their
wavefunctions are implicit.  That information is encoded in the
$\mathcal{B}$ coefficients.  To obtain them, one should
start with a microscopic spin-lattice coupling, expand it with
Holstein-Primakoff bosons, and then use the canonical Bogoliubov
transformation which achieves the diagonal form of Eq.~\eqref{eq:12}
to express the coupling as in Eq.~\eqref{eq:13}.
We apply this procedure to a particular case in Sec.~\ref{sec:square-lattice-two}. 
 The following general results hold beyond this specific case, and
 only assume Eqs.~(\ref{eq:12}, \ref{eq:13}) as a starting point.

To proceed to evaluate Eqs.~(\ref{eq:7}-\ref{eq:11}), we use Wick's
theorem (valid for the free boson Hamiltonian in Eq.~\eqref{eq:12}) to
compute the necessary correlation functions, which decomposes them
into products of the free-particle two-point function,
\begin{eqnarray}
\label{eq:15}
  \left\langle b^{q_1}_{\ell_1,\mb p_1,z_1}(t_1) b^{q_2}_{\ell_2,\mb
    p_2,z_2}(t_2)\right\rangle =
& \delta_{\ell_1,\ell_2}\delta_{z_1,z_2}\delta_{q_1,-q_2}\delta_{\mb p_1 + \mb p_2,\mb
  0}  \nonumber\\
 \times& f_{q_2}(\Omega_{\ell_1,q_1\mb p_1}) e^{-i q_2 \Omega_{\ell_2,q_2\mb p_2}(
  t_1 -t_2)}.\nonumber\\
\end{eqnarray}
Here $f_q(\Omega) = (1+q)/2 +n_{\rm B}(\Omega)$, where $n_{\rm B}(\Omega)$ is the Bose
distribution.
\begin{widetext}
This results in the following expressions for the diagonal and
off-diagonal scattering rates:
\begin{align}
  \label{eq:16}
  D^{(s)}_{n\mathbf{k}}=\frac{(3-s)\pi }{\hbar^2 N_{\rm uc}^{\rm 2d}} 
                         \sum_{\mb{p}} \sum_{\ell_1,\ell_2} \, \frac{\sinh(\tfrac
    \beta 2 \hbar\omega_{n\mathbf k})}
           {\sinh(\tfrac
    \beta 2 \hbar\Omega^{\ell,+}_{\mb p})\sinh(\tfrac
  \beta 2 \hbar\Omega^{\ell',-s}_{\mb p + \mb k})}  
\quad\delta(\omega_{n\mathbf k}-\Omega^{\ell,+}_{\mb p }-s
           \Omega^{\ell',-s}_{\mb p + \mb k})
     \left |\mathcal B^{n,\ell,\ell'|+s-}_{\mathbf k;\mb p+\frac{\mathbf{k}}{2}}
     \right |^2,
\end{align}
where $s=\pm$ and $D_{n\mb k} = \sum_s
  D^{(s)}_{n\mathbf{k}}+\breve D_{n\mathbf{k}}$, we defined
  $\Omega_{\mb p}^{\ell,q}=\Omega_{\ell,q\mb p}$, and for $q,q'=\pm1$ 
\begin{align}
\label{eq:17}
  \mathfrak{W}^{\ominus,qq'}_{n\mathbf{k},n'\mathbf{k}'}
   & = \frac{64\pi^2}{\hbar^4}\frac{1}{N_{\rm uc}^{2d}} \sum_{\mathbf{p}}\sum_{\{\ell_i,q_i\}} \mathfrak{D}_{q\mathbf{k}q'\mathbf{k}',\mathbf{p}}^{nn'|q_1q_2q_3,\ell_1\ell_2\ell_3}~\mathfrak{F}_{q\mathbf{k}q'\mathbf{k}',\mathbf{p}}^{q_1q_2q_4,\ell_1\ell_2\ell_3} ~ \mathfrak{Im} \Bigg\{ {\mathcal
    B}^{n\ell_2\ell_3|q_2q_3q}_{\mb k,\mb p +\frac{1}{2}q\mb k + q'\mb
    k'} {\mathcal
    B}^{n'\ell_3\ell_1|-q_3q_1q'}_{\mb k',\mb p+ \frac{1}{2}q'\mb k'}
    \nonumber \\
  & \times \textrm{PP} \Big[\frac{{\mathcal B}^{n\ell_1\ell_4|-q_1q_4 -q}_{\mb k,\mb p +
    \frac{1}{2}q\mb k} {\mathcal B}^{n'\ell_4\ell_2|-q_4-q_2-q'}_{\mb
    k',\mb p + q\mb k +\frac{1}{2}q'\mb k'}}{\Delta_{n\mathbf{k}n'\mathbf{k}'}^{qq'}+ q_1
    \Omega^{\ell_1,-q_1}_{\mb p}-q_2\Omega^{\ell_2,q_2}_{\mb p + q\mb k+q'\mb
    k'}-2q_4 \Omega^{\ell_4,-q_4}_{\mb p +q\mb k}}
    +\frac{{\mathcal B}^{n'\ell_1\ell_4|-q_1-q_4 -q'}_{\mb k',\mb p +
    \frac{1}{2}q'\mb k'} {\mathcal B}^{n\ell_4\ell_2|q_4-q_2-q}_{\mb
    k,\mb p + \frac{1}{2}q\mb k +q'\mb k'}}{\Delta_{n\mathbf{k}n'\mathbf{k}'}^{qq'}- q_1
    \Omega^{\ell_1,-q_1}_{\mb p}+q_2\Omega^{\ell_2,q_2}_{\mb p + q\mb k+q'\mb
    k'}-2q_4 \Omega^{\ell_4,q_4}_{\mb p +q'\mb k'}}\Big]\Bigg\},
\end{align}
where $\{\ell_i,q_i\}=\{\ell_1, \ell_2, \ell_3, \ell_4, q_1, q_2,
  q_3, q_4\}$ with $q_j=\pm1$ and $\ell_j$ runs over the magnon branch
indices, and where we used as before
$\Sigma^{q,q'}_{n\mathbf{k}n'\mathbf{k}'}=q\omega_{n\mathbf{k}}+q'\omega_{n'\mathbf{k}'}$,
$\Delta^{q,q'}_{n\mathbf{k}n'\mathbf{k}'}=q\omega_{n\mathbf{k}}-q'\omega_{n'\mathbf{k}'}$,
and we defined the product of delta functions $\mathfrak{D}$ and
`thermal factor' $\mathfrak{F}$:
\begin{eqnarray}
  \label{eq:24}
  \mathfrak{D}_{q\mathbf{k}q'\mathbf{k}',\mathbf{p}}^{nn'|q_1q_2q_3,\ell_1\ell_2\ell_3}
  &=&\delta\left(
    \Sigma_{n\mathbf{k}n'\mathbf{k}'}^{qq'}+q_1\Omega^{\ell_1,-q_1}_{\mb
    p}+q_2\Omega^{\ell_2,q_2}_{\mb p +q\mb k +  q'\mb  k'}\right) \delta
       \left(\Delta_{n\mathbf{k}n'\mathbf{k}'}^{qq'}+2q_3\Omega^{\ell_3,-q_3}_{\mb p+q'\mb  k'}
      -q_1\Omega^{\ell_1,-q_1}_{\mb   p}+q_2\Omega^{\ell_2,q_2}_{\mb   p+q\mb k +  q'\mb  k'}\right),\nonumber\\
  \mathfrak{F}_{q\mathbf{k}q'\mathbf{k}',\mathbf{p}}^{q_1q_2q_4,\ell_1\ell_2\ell_3}
  &=&q_4 \left(2n_{\rm B}(\Omega^{\ell_3,-q_3}_{\mb p+q'\mb k'})+1\right)
    \left(2n_{\rm B}(\Omega^{\ell_1,-q_1}_{\mb p})+q_1+1\right)\left(2n_{\rm B}(\Omega^{\ell_2,q_2}_{\mb p+q\mb k +q'\mb
    k'})+q_2+1\right) .
\end{eqnarray}
\end{widetext}
These formulae make no further assumptions on the nature of the spin
wave modes or spin-lattice couplings, and so could be applied to
general problems involving spin-lattice coupling using a spin wave
approach. Note that although we take the spin wave operators to be
  free bosons, with Gaussian correlations, the $Q_{n\mb{k}}^q$
  operator defined through Eq.~\eqref{eq:13} is generally
  non-Gaussian, as it is bilinear in the $b$ fields.

\subsection{Square lattice two-sublattice antiferromagnets}
\label{sec:square-lattice-two}

Now we evaluate the diagonal and Hall scattering rates specifically
for spin waves on the square lattice in low magnetic fields.  We
assume the magnon dispersions $\Omega_{\ell,\mb{k}}=\sqrt{v^2_{\rm
    m}\ul{k}^2+\Delta_\ell^2}$ ($\ell=0,1$ in this case) with magnon velocity $v_{\rm m}$ and magnon
gaps $\Delta_\ell$, and take isotropic acoustic phonons with
$\omega_{n\mb{k}} = v_{\rm ph} \sqrt{\ul{k}^2+k_z^2}$ (we define $\ul{k}=\sqrt{k_x^2+k_y^2}$).
We obtain the coefficients $\mathcal{B}_{\mb{k};\mb{p}}^{n\ell_1\ell_2|q_1q_2q}$
from the continuum description of the spin waves in terms of local
fluctuating uniform and staggered magnetization fields, and the 
symmetry-allowed couplings of these fields to the strain.  The
expressions for these coefficients are algebraically complicated and
some further details are given in Appendix~\ref{sec:deta-magn-model}, with a full exposition
of the calculations to be presented in a separate publication \cite{phononcompanion}.  Here
instead we sketch the important properties of the coefficients and
their origins.

\subsubsection{Scaling}
\label{sec:scaling}

First, when the temperature is smaller than the magnon gaps, $k_B T
\lesssim \Delta_\ell$, all contributions to scattering become
exponentially suppressed by thermal factors and the scaling is
unimportant.  For larger temperatures, the gaps are negligible, and
the momentum sum(s) in Eqs.~(\ref{eq:16},\ref{eq:17}) are
dominated by momenta of order $k,p \sim k_B T/v_{\rm m}$.  Then the
$\mathcal{B}$ coefficients, evaluated for momenta of this order, are
sums of three types of contributions:
\begin{equation}
  \label{eq:18}
  \mathcal{B} \sim \left(\frac{k_B T}{Mv_{\rm ph}^2}\right)^{\frac{1}{2}}n_0^{-1}
  \left(
    \lambda_{mm} \frac{\chi k_BT}{n_0} + \lambda_{mn} + \lambda_{nn} \frac{n_0}{\chi
      k_B T}\right).
\end{equation}
Here $M$ is the mass per unit cell of the solid, $n_0$ is the ordered
(staggered) moment density, $\chi$ is the uniform susceptibility, and
$\lambda_{mm},\lambda_{nn}$ and $\lambda_{mn}$ represent couplings of
the strain to exchange terms quadratic in local magnetization
fluctuations $\delta m$,
local staggered magnetization fluctuations $\delta n$, and the product of the
two, respectively.  Microscopically this arises from effects like
magnetostriction, the modification of orbital overlaps due to
strain-induced bond length and angle changes, etc.  The different powers of
temperature multiplying the different $\lambda$ couplings arise from
the fact that the order parameter of the antiferromagnet is the
staggered magnetization, and therefore its fluctuations are more
singular than those of the uniform magnetization, which is, however,
still a low energy mode in an antiferromagnet.

Depending upon the relative magnitudes of these different couplings,
distinct scalings are observed for the diagonal and off-diagonal scattering rates,
and hence for thermal conductivity components.  To
perform a full evaluation, we use parameters (given explicitly in Appendix~\ref{sec:deta-magn-model})
which describe a typical situation corresponding to weak spin-orbit
coupling and correspondingly weak anisotropy of magnetic exchange.  In
this case, there is a hierarchy that $\lambda_{mm} \gg \lambda_{nn},\lambda_{mn}$
(which is ultimately a consequence of Goldstone's theorem).
Furthermore, in the low field regime, i.e.\ when the field-induced
magnetization $m_0$ of the antiferromagnet is much smaller than $m_s$ the
saturation value, $m_0 \ll m_s$, the mixed coupling $\lambda_{mn}$ is
proportional to $m_0$ and $\lambda_{mn} \ll \lambda_{nn}$ as well.

These facts allow one to estimate the scalings of the important
physical quantities.  The longitudinal scattering rate
(Eq.~\eqref{eq:16}) scales as
\begin{equation}
  \label{eq:19}
 D_{n\mb{k}} \sim  \frac{1}{\tau} \sim T^{d-1} |\mathcal{B}|^2 \sim T^{d+2x}.
\end{equation}
Here $d$ is the dimensionality of the spin system (which we take later
equal to $d=2$ for numerical calculations) while phonons are always
three dimensional.  The crucial exponent $x=1$ occurs in the ``high''
temperature regime dominated by $\lambda_{mm}$, while a crossover
to behavior controlled by $\lambda_{nn}$ with $x=-1$ can occur at
lower temperature if the minimum magnon gap is sufficiently small.
This behavior corresponds to the longitudinal thermal conductivity (Eq.~\eqref{eq:25})
behaving as
\begin{equation}
  \label{eq:20}
  \kappa_L \sim T^{3-d-2x},
\end{equation}
when magnon-phonon scattering dominates the phonon mean free path. 
Again the power laws apply in certain distinct regimes, and should be
pieced together, along with the influence of non-zero gaps and other
scattering mechanisms of phonons, to form a complete picture of the
thermal conductivity.  This is captured in the numerical calculations.

Next, we turn to the thermal Hall effect.  It is crucial to keep in
mind the {\em effective} time-reversal symmetry of an antiferromagnet
under the combined action of time-reversal and a translation which
exchanges the two sublattices.  The uniform magnetization is invariant
under this symmetry but the staggered magnetization is odd.
Consequently, the couplings $\lambda_{mm}$ and $\lambda_{nn}$ are even
under effective time-reversal, while only $\lambda_{mn}$ is odd.  This
implies that the Hall conductivity and Hall
scattering rate $\mathfrak{W}^{\ominus,{\rm eff}}$, with $\mf W^{\ominus,{\rm eff},qq'}_{n\mb k,n'\mb k'} := 
\mf W^{\ominus,qq'}_{n\mb k,n'\mb k'}+\mf W^{\ominus,qq'}_{n-\mb k,n'-\mb k'}$, which are odd under effective
time-reversal, must be proportional to an odd power of
$\lambda_{mn}$, and to linear order in the magnetic
field/average magnetization, these quantities are simply linear in $\lambda_{mn}$.  From
Eq.~\eqref{eq:17} and Eq.~\eqref{eq:18}, we therefore obtain
\begin{equation}
  \label{eq:21}
  \mathfrak{W}^{\ominus,\rm eff} \sim T^{d-1} \lambda_{mn} \left(
    \lambda_{mm} T + \lambda_{nn} T^{-1}\right)^3 \sim T^{d-1+3x}.
\end{equation}
The natural definition of a skew scattering rate multiplies the above
by a phase-space factor to account for the sum over different final
states of the scattering, which gives $1/\tau_{\rm skew}
\sim  T^3 \mathfrak{W}^{\ominus,\rm eff}  \sim T^{d+2+3x}$.  

We would like to emphasize that within any scattering mechanism of
phonon thermal Hall effect, the skew scattering rate is a more fundamental
measure of the chirality of the phonons than the thermal Hall {\em
  conductivity}.  This is because the Hall conductivity inevitably
involves the combination of the skew and longitudinal scattering rates
(in the form $\tau^2/\tau_{\rm skew}$), and the longitudinal
scattering rate of phonons has many other contributions that do not
probe chirality, and may have complex dependence on temperature and
other parameters that obscure the skew scattering.

Consequently, instead of the thermal Hall conductivity we will discuss
the thermal Hall resistivity, $\varrho_H$, which is simply
proportional to $1/\tau_{\rm skew}$,
at least in the simplest view where the angle-dependence of the
longitudinal scattering does not spoil its cancellation. 

We define the thermal Hall resistivity tensor as usual by the matrix inverse,
$\bs{\varrho} = \bs{\kappa}^{-1}$.
In particular, considering the simplest case of isotropic
$\kappa^{\mu\mu}\rightarrow \kappa_L$ and $\kappa_{L} \gg
\kappa^{\mu\neq\nu}$, one thus has
\begin{equation}
\label{eq:22}
  \varrho^{\mu\nu}_H = \frac {\varrho_{\mu\nu}-\varrho_{\nu\mu}} 2
  \approx \frac{-\kappa_{\mu\nu}+\kappa_{\nu\mu}}{2\kappa_{L}^2}
  = - \frac{\kappa^{\mu\nu}_{H}}{\kappa_{L}^2}.
\end{equation}
The quantity $\varrho^{\mu\nu}_{H}$ is independent of the scale of the
longitudinal scattering, in the sense that under a rescaling
$D_{n\mb{k}}\rightarrow \zeta D_{n\mb{k}}$, then
$\varrho^{\mu\nu}_{H}$ is unchanged.  If we assume that $D_{n\mb{k}} = 1/\tau$ is
$(n,\mb{k})$-independent, e.g.\ as is the case if dominated by some
extrinsic effects, then we can readily extract the  scaling of the
thermal Hall resistivity.  One finds
\begin{equation}
\label{eq:23}
  \varrho_H \sim \mf W^{\ominus,\rm eff} \sim T^{d-1+3x},
\end{equation}
which is verified numerically.  This scaling behavior should also be
roughly true in the presence of more angle-dependent scattering, given
the aforementioned independence on the scale of scattering.

Finally, we comment on the role of spin-orbit coupling in the
  present model. The coefficient $\lambda_{mn}$ communicates the lack
  of effective time-reversal and mirror symmetry breaking to the
  scattering rate $\mathfrak{W}^\ominus$, and thereby the Hall
  resistivity begins at linear order in this coefficient. In the
  present model, $\lambda_{mn}$ is also proportional to (symmetric)
  spin-orbit coupling terms (microscopically, derivatives of such
  terms with respect to ionic displacement)---see Appendix~\ref{sec:deta-magn-model}.  In general, however, for more complex
  magnetic ordering patterns, a nonzero Hall effect may be obtained
  from our formulation even in the absence of spin-orbit coupling.

\subsubsection{Numerical evaluation}
\label{sec:numerical-evaluation}

It is important to verify that the formulae in
Eqs.~(\ref{eq:16},\ref{eq:17}) are sufficient to generate all the
expected symmetry-allowed scattering processes and thereby
contributions to the thermal Hall conductivity.  To do so, we
evaluated these formulae numerically, which also allows a test of the
scaling predictions above.  In the numerical calculations, we took
specific values for the microscopic parameters which define the
dispersions of the magnons and phonons, as well as those which underlie
the $\mathcal{B}$ coefficients, comprising spin-lattice couplings and
the mass per unit cell.
Eqs.~(\ref{eq:16},\ref{eq:17},\ref{eq:25},\ref{eq:7}) were evaluated by a C
code using the Cuba and Cubature libraries for numerical integration \cite{cubapaper}.

It is convenient to measure energies in units of the phononic energy scale
$\epsilon_0=k_B T_0=\hbar v_{\rm ph}/\mathfrak{a}$, which is equal to the Debye temperature up to a factor, and we report
thermal conductivities in units of
$\kappa_0= k_B v_{\rm ph}/\mathfrak{a}^2$, which gives a natural scale
for phononic heat transport.

To make our numerical calculations more directly relevant, we loosely
chose key dimensionless parameters to loosely match those of Copper
Deuteroformate Tetradeuterate (CFTD), a square lattice S=1/2
antiferromagnet which has been intensively studied via neutron
scattering
\cite{christensen2007quantum,dalla2015fractional,ronnow2001spin} due
to its convenient scale of exchange which suits such measurements. For
our purposes, CFTD has the desirable attribute that the magnon and
phonon velocities are comparable (based on an estimate of the sound
velocity from the corresponding hydrate \cite{kameyama1973elastic}),
which creates a significant phase space for magnon-phonon scattering.
In particular, we take $v_{\rm m}/v_{\rm ph} = 2.5$, while the
corresponding ratio in La$_2$CuO$_4$, $v_{\rm m}/v_{\rm ph}$ is approximately 30.
We also use the mass per unit cell $M_{\rm uc}$ appropriate to CFTD.
The parameters $n_0=1/2$ and $\chi = 1/(4J \mathfrak{a}^2)$ are chosen
to be consistent with spin wave theory.  We included small magnon
gaps, $\Delta_0 = 0.2 \epsilon_0$ and $\Delta_1 = 0.04 \epsilon_0$.  The
microscopic spin-lattice couplings were taken consistent with the
expectations for weak spin-orbit coupling, and are given in Appendix~\ref{sec:deta-magn-model}.  Finally, we included in the calculations a small constant
contribution $\breve{D}_{n\mb k}\rightarrow\gamma_{\rm ext}$,
independent of $(n,\mb k)$, to model additional scattering channels.  In very
clean monocrystals and in the absence of any other phonon scattering
events, $\gamma_{\rm ext}\sim v_{\rm ph}/L$ reduces to the rate at
which phonons bounce off the boundaries of the sample (of size $L$).
We vary $\gamma_{\rm ext}$ to show the dependence on these extrinsic
effects.  For the calculations of the Hall effect we include a small
non-zero magnetization in the direction of the applied field, of
1/20th of the saturation magnetization.

\begin{figure*}[htbp]
  (a)
\begin{subfigure}[b]{0.45\textwidth}
  \includegraphics[width=\columnwidth]{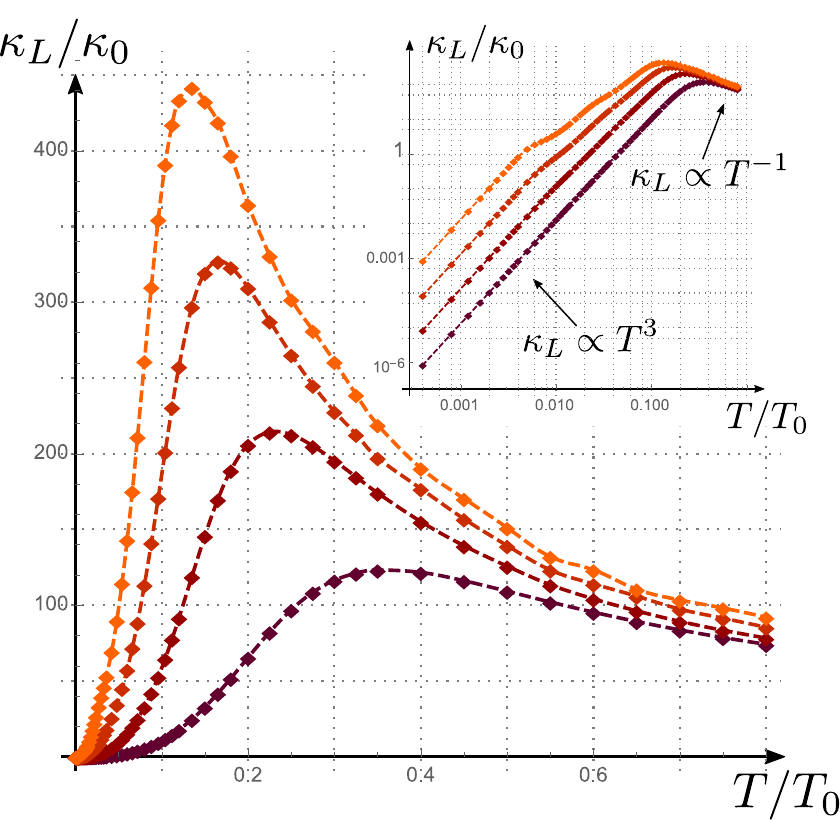}
\end{subfigure}
\hfill
(b)
\begin{subfigure}[b]{0.45\textwidth}
  \includegraphics[width=\columnwidth]{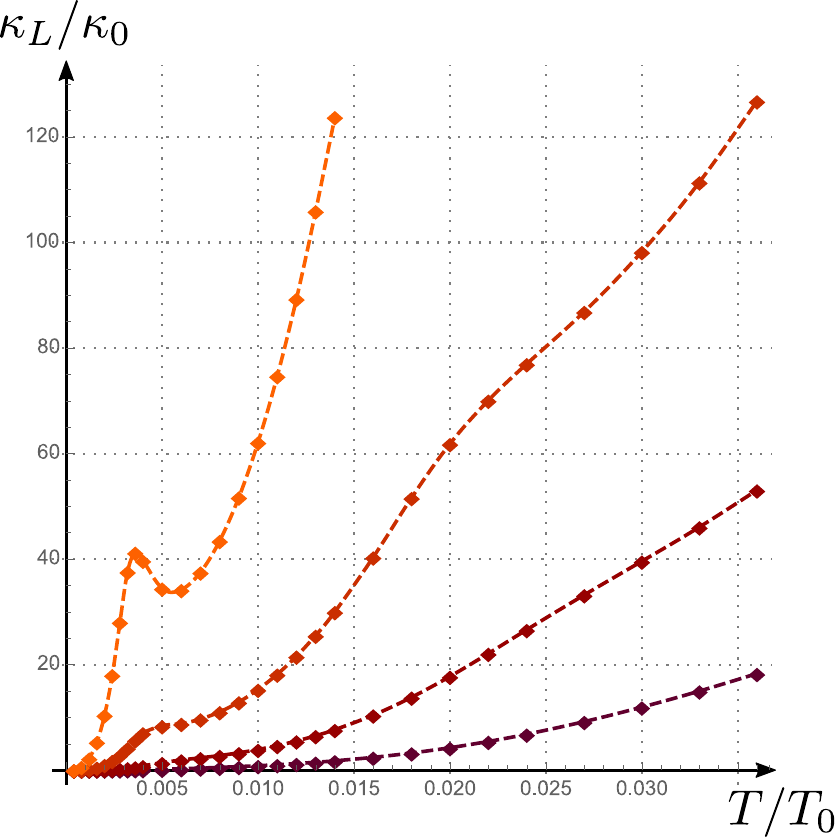}
  \end{subfigure}
  \caption{Longitudinal thermal conductivity $\kappa_L$ with respect to temperature $T$,
    for four different values of 
    $\gamma_{\rm ext}$. (a) Results on an order one temperature scale,
    with $\gamma_{\rm ext}= 1\cdot 10^{-z}(v_{\rm ph}/\mathfrak{a}), z \in \llbracket 4, 7\rrbracket$,
    from darker $(z=4)$ to lighter $(z=7)$ shade. A crossover occurs between two scaling regimes with
    $x=1$ and $x=-1$ (see Eq.~\eqref{eq:20}). Inset: log-log plot; the scaling behaviors are consistent with the analysis
  presented in the text. (b) Results on a smaller temperature scale, with $\gamma_{\rm ext}= 1\cdot 10^{-z}(v_{\rm ph}/\mathfrak{a}), z \in \llbracket 6, 9\rrbracket$,
    from darker $(z=6)$ to lighter $(z=9)$ shade. The peaks are
    features related to the magnon gaps.}
     \label{fig:kappaLplots}
   \end{figure*}

Fig.~\ref{fig:kappaLplots} shows the results for the longitudinal
thermal conductivity versus temperature in zero or low applied
magnetic field (the results are insensitive to small magnetizations),
for different choices of $\gamma_{\rm ext}$.  In panel (a), a broad
temperature range is shown, which exposes the evolution from an
extrinsic scattering regime $\kappa_L \propto T^3$ at low temperature
to an intrinsic one $\kappa_L \propto 1/T$ at high temperature.
In panel (b), further features emerge related to the scales of the magnon gaps.

\begin{figure}[htbp]
  \centering
  \includegraphics[width=0.9\columnwidth]{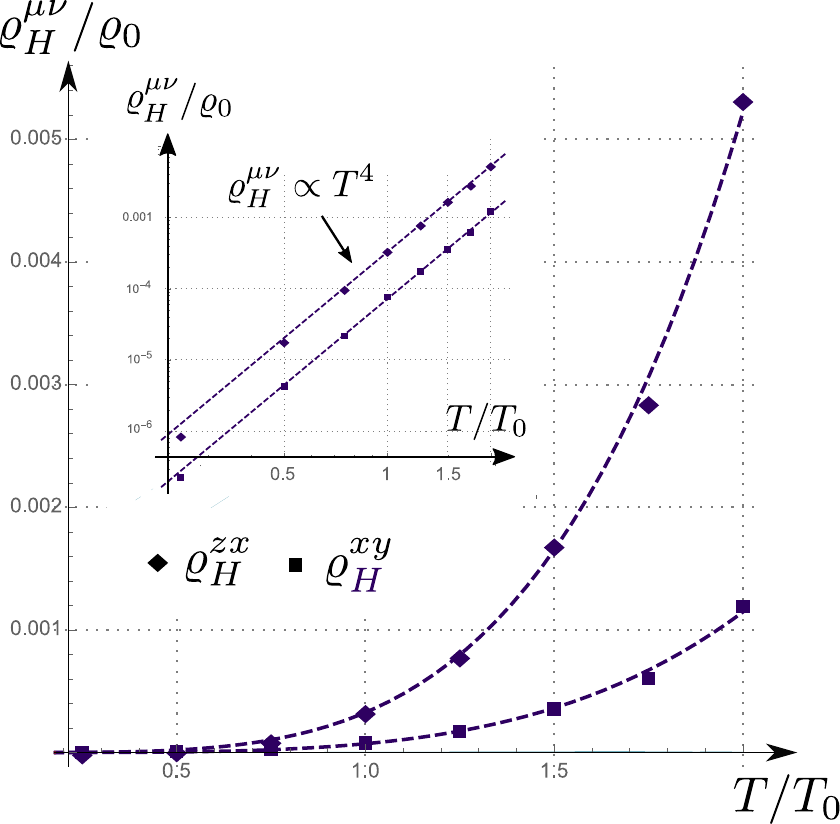}
  \caption{Thermal Hall resistivity $\varrho_H^{xy}$ and $\varrho_H^{zx}$ (in units of $\varrho_0=\kappa_0^{-1}$ with respect to temperature $T$. The transverse magnetization values $(m_0^y, m_0^z)$ used for computing $\varrho_H^{xy}$ and $\varrho_H^{xz}$ were $(\mt {0.0}, \mt{0.05})$ and $(\mt {0.05}, \mt{0.0})$, respectively. Inset: log-log plot; the scaling behavior is consistent with the analysis presented in the text.}
  \label{fig:rhoH}
\end{figure}

Next we turn to the calculations of chiral scattering and the thermal
Hall effect.  Figure~\ref{fig:WH} shows a density plot of the Hall
scattering rate $\mathfrak{W}^{\ominus,-+}$ as a function of the polar
angle of $\mathbf{k}'$, $\theta(\mb k')$, and of the difference in
azimuthal angles of $\mathbf{k}'$ and $\mathbf{k}$, $\varphi(\mb k,\mb k')=\phi(\mathbf{k}')-\phi(\mathbf{k})$.  We see that it
has an intricate structure reflecting kinematics and energetics.  The
thermal Hall resistivity in the constant longitudinal scattering
approximation ($D_{n\mb{k}} = 1/\tau$ independent of $n,\mb{k}$) is
plotted in Fig.~\ref{fig:rhoH} for two different field orientations:
for a field along the $z$ axis, normal to the planes, we plot
$\varrho_H^{xy}$, and for a field along the $y$ axis, within the
planes, we plot $\varrho_H^{zx}$.  Both curves perfectly fit the $T^4$
scaling expected theoretically from Eq.~\eqref{eq:23} (using
$d=2,x=1$) with weak spin-orbit coupling.  Notably the magnitudes of
the thermal Hall resistivity for the two orientations are comparable,
and it is actually {\em larger} for an {\em in-plane} field than for
an out-of-plane one!  

\section{Conclusion}
\label{sec:conclusion}

In this paper, we presented a theory for the skew scattering of phonons
coupled to a quantum collective field, which gives rise to a phonon
thermal Hall effect.  A general formula, given in
Eqs.~(\ref{eq:7},\ref{eq:8}), allows the latter to be calculated for
{\em arbitrary} correlations of the collective variable.  We then explicitly
calculated these correlations for the case in which the
collective field is bilinear in canonical bosons, e.g.\ spin wave
operators.  A formula with no further assumptions is given in
Eq.~\eqref{eq:17}.  Applying this to the regime of long-wavelength
magnons in a square lattice antiferromagnet, we obtained a non-zero
thermal Hall effect and its scaling with temperature in various
regimes.

While we are not aware of any general results on the intrinsic phonon
Hall conductivity due to scattering, there are a number of
complementary theoretical papers as well as some prior work which
overlap a small part of our results.  The specific problem of phonons
scattering from magnons was studied long ago to the leading second
order in the coupling by Cottam \cite{cottam_spin-phonon_1974}.  That
work, which assumed the isotropic SU(2) invariant limit, agrees with
our calculations when these assumptions are imposed.  The
complementary mechanism of intrinsic phonon Hall effect due to phonon
Berry curvature was studied by many authors
\cite{qin2012berry,saito2019berry,zhang2010topological,zhang2021phonon},
including how the phonon Berry curvature is induced by spin-lattice
coupling in Ref.~\cite{ye2021phonon}.  The majority of recent
theoretical work has concentrated on {\em extrinsic} effects due to
scattering of phonons by
defects \cite{sun2021large,guo2021extrinsic,guo2022resonant,flebus2021charged}.
The pioneering paper of Mori {\em et al.\ }\cite{mori} in particular
recognized the importance of higher order contributions to scattering
for the Hall effect, and is in some ways a predecessor to our work.

Do the present results explain experiments on the cuprates?  We have
not attempted a quantitative comparison, for several reasons. This
would require some detailed knowledge of spin-lattice couplings.  It
also is numerically difficult because in the cuprates there is a very large
ratio of magnon to phonon velocities (of order $30$), which renders
the scattering phase space narrow and the integration challenging.
Nevertheless, it is interesting to ask about the order of magnitude of
the response.  For this comparison, we  follow the logic outlined in
Sec.~\ref{sec:scaling} in which we argued that the thermal Hall
resistivity is a better quantity for which to compare theory and
experiment.  We obtain the thermal resistivity from the experimental data in
Ref.~\cite{grissonnanche2020chiral} on the undoped material La$_2$CuO$_4$: at about 20~K, the longitudinal
conductivity $\kappa_{xx} \approx 10$~W/(K m) (from their extended
data Figure 2), and the thermal Hall conductivity $\kappa_{xy} \approx
40$~mW/(K m).  Using Eq.~\eqref{eq:22} and the the value
$\varrho_0^{\rm LCO}
\approx 2.6$~K m/W, we then obtain $(\varrho_H/\varrho_0)^{\rm LCO} \approx
1.5\times 10^{-4}$.    This is at least comparable to values in Figure~\ref{fig:rhoH}.

Regardless of whether the intrinsic picture is correct for the
cuprates (we think it most promising for systems like CFTD for which
there is a good phase space match of phonons and magnons), we believe
that a scattering mechanism {\em of some kind} is very likely at
work.  Therefore, we would encourage analysis of future experimental data in
terms of $\varrho_H$ rather than $\kappa_H$.  

In a companion paper \cite{phononcompanion}, we will expound on the results of the present
paper and give several extensions covering even more general types of
coupling of phonons to collective degrees of freedom.  There also
remain many other related problems that would be interesting to
explore, for example the influence of electronic disequilibrium upon
the phonons, and vice versa, and the interplay of scattering, presumed
here to be dominant, and phononic Berry phases.  We hope that the
present study provides a theoretical framework to begin to approach
these and other intriguing questions.

\acknowledgements

We thank Mengxing Ye for
valuable discussions, as well as Xiao Chen and Jason Iaconis for a
collaboration on a related topic. We also sincerely acknowledge Roser
Valent\'i for her encouragements and enthusiasm. The premises
of this project were funded by the Agence Nationale de la Recherche through Grant ANR-18-ERC2-0003-01 (QUANTEM). The bulk of this project was funded by the European Research Council (ERC) under the European Union’s Horizon 2020 research and innovation program (Grant agreement No.\ 853116, acronym TRANSPORT). L.B.\ was supported by the DOE, Office of Science, Basic Energy Sciences under Award No.\ DE-FG02-08ER46524.  It befits us to acknowledge the hospitality of
the KITP, where part of this project was carried out, funded under NSF
Grant NSF PHY-1748958.

\bibliography{phononpaper.bib}

\appendix

\section{From interaction terms to the collision integral}
\label{sec:from-inter-terms}

\subsection{First Born order}
\label{sec:first-born-order}

First, we consider only the first term of Born's
expansion. The transition rate associated with $H'$
at this order involves the matrix elements:
\begin{equation}
\label{eq:app14}
    T^{[1]}_{\mathtt{i}\rightarrow\mathtt{f}}    = \sum_{n\mathbf{k}q}\sqrt{N^i_{\mathbf{k},n}+\tfrac{q+1}2}~\langle
    f_s|Q^{q}_{n\mathbf k}|i_s\rangle~ \mathbb{I}({i_p}\overset{q\cdot n\mathbf k}{\longrightarrow}{f_p}),
  \end{equation}
where $\mathbb{I}({ i_p}\overset{q\cdot n\mathbf k}{\longrightarrow}{
  f_p})$ means that the only difference between $| i_p \rangle$ and $|
f_p\rangle$ is that there is $q=\pm1$ more phonon of species
$(n,\mathbf k)$ in the final state.

We then compute the squared matrix element. We have 
\begin{eqnarray}
\label{eq:app15}
  \left|T^{[1]}_{\mathtt{i}\rightarrow\mathtt{f}}\right|^2 &=& 
\sum_{n\mathbf{k}q}\left(N^i_{\mathbf{k},n}+\frac{q+1}{2}\right)~\mathbb{I}({i_p}\overset{q\cdot n\mathbf k}{\longrightarrow}{f_p})\nonumber\\
&&\times \quad\langle
i_s|Q^{-q}_{n\mathbf k}|f_s\rangle\langle
f_s|Q^{q}_{n\mathbf k}|i_s\rangle.
\end{eqnarray}
We then enforce the energy conservation
$\delta(E_{\mathtt{f}}-E_{\mathtt{i}}) =
\delta(q\omega_{n\mathbf{k}}+E_{f_s}-E_{i_s})$ by writing the latter
as a time integral, i.e.\ $\int_{-\infty}^{+\infty}{\rm d}t e^{i\omega
  t}=2\pi\delta(\omega)$, identify $A(t)=e^{+iHt}Ae^{-iHt}$, use the
identity $1=\sum_{ f_s}| f_s\rangle\langle f_s|$, and take the $Q$ field
in the initial state to be in thermal equilibrium $p_{
  i_s}=Z_s^{-1}e^{-\beta E_{ i_s}}$. Finally summing over $| i_s\rangle$
and identifying $\langle A \rangle_\beta = Z_s^{-1}\text{Tr}(e^{-\beta
  H}A)$, we find that the scattering rate between phonon states at first
Born's order reads
\begin{eqnarray}
\label{eq:app17}
  \Gamma^{[1];[1]}_{ i_p \rightarrow  f_p}&=& 
\sum_{n\mathbf k q}
                                              \left (N^i_{n\mathbf{k}}
                                              + \tfrac{q+1}{2}\right )
  ~\mathbb{I}({ i_p}\overset{q\cdot n\mathbf k}{\longrightarrow}{ f_p})\\
  &&\times ~
\int_{-\infty}^\infty \! \text dt\, e^{-i q\omega_{n\mathbf{k}} t}  \left\langle
Q^{-q}_{n\mathbf k}(t)~Q^{q}_{n\mathbf k}(0)\right\rangle_\beta.\nonumber
\end{eqnarray}
To arrive at the collision integral, the final step involves summing over final phononic states $ f_p$ and
taking the average over initial phononic states $i_p$. First, we notice that a change of variables in $\left\langle
Q^{-q}_{n\mathbf k}(t)~Q^{q}_{n\mathbf k}(0)\right\rangle_\beta$ leads to the detailed-balance relation
\begin{equation}
\label{eq:app28}
\left\langle
Q^{-q}_{n\mathbf k}(t)~Q^{q}_{n\mathbf k}(0)\right\rangle_\beta= \left\langle
Q^{q}_{n\mathbf k}(t)~Q^{-q}_{n\mathbf k}(0)\right\rangle_\beta e^{-q\beta\omega_{n\mathbf k}}.
\end{equation}
It is then straightforward to show that only the commutator term on
the right-hand-side of Eq.~\eqref{eq:app17} satisfies this
relation. In turn, the final expression for the diagonal of the
collision matrix takes the form of the spectral function:
\begin{equation}
\label{eq:app30}
 D^{[1];[1]}_{n\mathbf k} = - \int_{-\infty}^{+\infty}\text dt e^{-i\omega_{n\mathbf k}t}\left \langle [Q^{-}_{n\mathbf k}(t),
Q^+_{n\mathbf k}(0)]\right\rangle_\beta,
\end{equation}
as quoted in the main text.

\subsection{Second Born order}
\label{sec:second-born-order}

As discussed, the first Born approximation alone does not
lead to a nonzero thermal Hall effect. Here we compute that which
appears when the Born expansion is taken up to the second Born
order. We have
\begin{eqnarray}
\label{eq:app23}
T^{[1,1]}_{\mathtt{i}\rightarrow\mathtt{f}} &=& 
\sum_{n\mathbf{k},n'\mathbf{k}'}\sum_{q,q'=\pm}\sqrt{N^i_{n\mathbf{k}}+\tfrac{1+q}{2}}\sqrt{N^f_{n'\mathbf{k}'}+\tfrac{1-q'}{2}}\nonumber\\
&&\cdot\sum_{ m_s}\frac{\langle
    f_s|Q^{q'}_{n'\mathbf k'}|m_s\rangle\langle
   m_s|Q^{q}_{n\mathbf k}|i_s\rangle}{E_{i_s}-E_{m_s}-q\omega_{\mathbf{k},n}+i\eta}
   ~\mathbb{I}({ i_p}\overset{q\cdot n\mathbf k}{\underset{q'\cdot n'\mathbf k'}\longrightarrow}{ f_p}),\nonumber\\
    &&
\end{eqnarray}
The squared $T$-matrix elements now include
cross-terms between the first and second orders of Born's
expansion. Here we give details of the calculation of one term, the
square of Eq.~\eqref{eq:app23},
$\left|T^{[1,1]}_{\mathtt{i}\rightarrow\mathtt{f}}\right|^2$. In the numerator, the matrix elements of the $Q$ operators can combine themselves in two
different ways, which we denote in the following as $(a)$: $\langle
    i_s|Q^{q}_{n\mathbf k}|m_s\rangle\langle
    m_s|Q^{q'}_{n'\mathbf k'}|f_s\rangle \langle
    f_s|Q^{-q'}_{n'\mathbf k'}|m'_s\rangle\langle
    m'_s|Q^{-q}_{n\mathbf k}|i_s\rangle$, and $(b)$: $\langle
    i_s|Q^{q}_{n\mathbf k}|m_s\rangle\langle
    m_s|Q^{q'}_{n'\mathbf k'}|f_s\rangle \langle
    f_s|Q^{-q}_{n\mathbf k}|m'_s\rangle\langle
    m'_s|Q^{-q'}_{n'\mathbf k'}|i_s\rangle$.

We use the following time integral representation of each of the denominators (using a regularized definition of the sign function),
\begin{eqnarray}
\label{eq:app22}
    \frac 1 {x \pm i\eta} &=& {\rm PP} \frac 1 x \mp i\pi \delta(x)\\
    &=& \frac 1 {2i} \int_{-\infty}^{+\infty}\text d t_1 e^{it_1 x}\text{sign}(t_1)\pm \frac 1 {2i} \int_{-\infty}^{+\infty}\text d t_1 e^{it_1 x}.\nonumber
\end{eqnarray}
and a introduce a third time integral to enforce the energy conservation $
E_{\mathtt{f}}-E_{\mathtt{i}}=q'\omega_{n'\mathbf{k}'}+q\omega_{n\mathbf{k}}+E_{f_s}-E_{i_s}$.
The product of the denominators (cf.\ Eq.~\eqref{eq:app22}) leads to
four terms, which can be labeled by two signs $s,s'=\pm$, and we
define, for convenience,
\begin{equation}
\Theta_{ss'}(t_1,t_2):=\left[-\text{sign}(t_1)\right]^{\frac{1-s}{2}}\left[\text{sign}(t_2)\right]^{\frac{1-s'}{2}}.
\end{equation}
\begin{widetext}
Then, the transition rate coming from this part of the total squared matrix
element can be written as a sum of eight terms:
\begin{eqnarray}
\label{eq:app24}
  \Gamma^{[1,1];[1,1]}_{ i_p \rightarrow  f_p}
  &=& \sum_{n\mathbf k, n'\mathbf k'}\sum_{q,q'}
      \left(N_{n\mathbf{k}}^i+\tfrac{q+1}2\right) \left(N_{n'\mathbf{k}'}^i+\tfrac{q'+1}2\right)
      \cdot\sum_{s,s'=\pm} \sum_{i=a,b}W^{[1,1];[1,1],(i),ss'}_{n\mathbf{k}q,n'\mathbf{k}'q'}
         ~\mathbb{I}({ i_p}\overset{q\cdot n\mathbf k}{\underset{q'\cdot n'\mathbf k'}\longrightarrow}{ f_p}),
\end{eqnarray}
where we defined (notice the order of the
first two operators in the correlator and the sign $t_1\pm t_2$ in the exponential):
\begin{eqnarray}
\label{eq:app25}
    W^{[1,1];[1,1],{(a)},ss'}_{n\mathbf{k}q,n'\mathbf{k}'q'} &=& \int \text d t\text d t_1 \text d t_2 \Theta_{ss'}(t_1,t_2) e^{i(q\omega_{n\mathbf{k}}+q'\omega_{n'\mathbf{k}'})t} e^{i(t_1+t_2) (q\omega_{n\mathbf{k}}-q'\omega_{n'\mathbf{k}'})}\nonumber\\
    &&\cdot \left\langle Q_{n\mathbf k}^{-q}(-t-t_2)Q_{n'\mathbf k'}^{-q'}(-t+t_2)Q_{n'\mathbf k'}^{q'}(-t_1)Q_{n\mathbf k}^{q}(+t_1)\right\rangle_\beta\\
     W^{[1,1];[1,1],{(b)},ss'}_{n\mathbf{k}q,n'\mathbf{k}'q'} &=& \int \text d t\text d t_1 \text d t_2 \Theta_{ss'}(t_1,t_2) e^{i(q\omega_{n\mathbf{k}}+q'\omega_{n'\mathbf{k}'})t} e^{i(t_1-t_2) (q\omega_{n\mathbf{k}}-q'\omega_{n'\mathbf{k}'})}\nonumber\\
     &&\cdot \left\langle Q_{n'\mathbf k'}^{-q'}(-t-t_2)Q_{n\mathbf k}^{-q}(-t+t_2)Q_{n'\mathbf k'}^{q'}(-t_1)Q_{n\mathbf k}^{q}(+t_1)\right\rangle_\beta.
\end{eqnarray}
\end{widetext}
One can 
show the following (``anti-'')detailed-balance relations
\begin{equation}
\label{eq:app37}
W^{[1,1];[1,1],(a),ss'}_{n\mathbf{k}q,n'\mathbf{k}'q'} = ss'~W^{[1,1];[1,1],(a),s's}_{n'\mathbf{k}'-q',n\mathbf{k}-q}
e^{-\beta(q\omega_{n\mathbf{k}}+q'\omega_{n\mathbf{k}'})},
 \end{equation}
 \begin{equation}
 \label{eq:app371}
 W^{[1,1];[1,1],(b),ss'}_{n\mathbf{k}q,n'\mathbf{k}'q'} = ss'~W^{[1,1];[1,1],(b),s's}_{n\mathbf{k}-q,n'\mathbf{k}'-q'}
 e^{-\beta(q\omega_{n\mathbf{k}}+q'\omega_{n'\mathbf{k}'})}.
\end{equation}
From this, the same holds for the symmetrized in $n\mathbf
k q\leftrightarrow n'\mathbf k' q'$ scattering rate
$\mathcal W^{[1,1];[1,1],ss'}_{n\mathbf kq,n'\mathbf k'q'}=\sum_{i=a,b}  W^{[1,1];[1,1],(i),ss'}_{n\mathbf k q,n'\mathbf k'q'} 
 + (n\mathbf k q\leftrightarrow n'\mathbf k'q')$, i.e.
\begin{equation}
\label{eq:app38}
\mathcal W^{[1,1];[1,1],ss'}_{n\mathbf k q,n'\mathbf k'q'} = ss'~e^{-\beta(q\omega_{n\mathbf{k}}+q'\omega_{n'\mathbf{k}'})}
\mathcal W^{[1,1];[1,1],ss'}_{n\mathbf k-q,n'\mathbf k'-q'}.
\end{equation}
We can then identify
\begin{eqnarray}
\label{eq:app39}
  \mathfrak{W}^{\ominus, [1,1];[1,1],qq'}_{n\mathbf{k},n'\mathbf{k}'}
  &=& N_{\rm uc}\sum_{s=\pm}  \mathcal{W}^{[1,1];[1,1],s,-s}_{n\mathbf{k}q,n'\mathbf{k}'q'} ,\\
\mathfrak{W}^{\oplus, [1,1];[1,1],qq'}_{n\mathbf{k},n'\mathbf{k}'}
   &=& N_{\rm uc}\sum_{s=\pm}  \mathcal{W}^{ [1,1];[1,1],ss}_{n\mathbf{k}q,n'\mathbf{k}'q'}, 
\end{eqnarray}
which, by construction, satisfy 
\begin{equation}
  \label{eq:41}
  \mf W^{\sigma, [1,1];[1,1],qq'}_{n\mathbf{k},n'\mathbf{k}'} =
  \sigma~e^{-\beta(q\omega_{n\mathbf{k}}+q'\omega_{n'\mathbf{k}'})}
  \mf W^{\sigma, [1,1];[1,1],-q-q'}_{n\mathbf{k},n'\mathbf{k}'},
\end{equation}
where $\sigma=\oplus$ (resp.\ $\sigma=\ominus$) indicates
that $\mf W$ satisfies detailed balance (resp.\ ``anti-detailed
balance''). Only $\mf W^{\ominus,
  [1,1];[1,1],qq'}_{n\mathbf{k},n'\mathbf{k}'}$ contributes to the
thermal Hall conductivity.

\section{Details of the magnetic model}
\label{sec:deta-magn-model}

\subsection{General symmetry-allowed model}
\label{sec:gener-symm-allow}

We begin with a semi-microscopic coupling of the local strain tensor $\bs {\mathcal{E}}_{\mb{r}}$
  to continuum non-linear sigma model fields: the density ${\rm m}_a$ of uniform
magnetization and ${\rm n}_a$ of staggered magnetization ($a=x,y,z$).  This is
\begin{align}
  \label{eq:30}
   & \mathcal{H}'_{\rm tetra}(\mb{r})
    =\\
  & \sum_{\substack{\alpha,\beta=x,y,z \\a,b=x,y,z}}\!\!\!\mathcal{E}^{\alpha\beta}_{\mb{r}}\left.\left(\Lambda_{ab}^{({\rm
      n}),\alpha\beta}{\rm n}_a{\rm  n}_b+\frac{\Lambda_{ab}^{({\rm
        m}),\alpha\beta}}{n_0^2}{\rm
      m}_a{\rm m}_b\right)\right|_{\mb{x},z},\nonumber
\end{align}
where $n_0$ is the ordered moment density.  The
$\Lambda^{(\xi),\alpha\beta}_{ab}$ coefficients are constrained by the
tetragonal symmetry of the crystal.  The non-linear sigma model is
defined by the constraints $\mb n\cdot \mb m = 0$ and $\mb n^2+\mb
m^2/n_0^2=1$.

We expand the above to second
order in the {\em fluctuations} $(\delta m,\delta n)$ around the average values due
to both spontaneous ordering and the applied field.   We take the
N\'eel vector along $\hat{\bs x}$.  Then ${\rm n}_x = 1 - \frac 1 2
\sum_{b=y,z}[\delta n_b^2+\frac 1 {n_0^2}(m_0^b+\delta m_b)^2]$, and $
{\rm m}_x = -\sum_{b=y,z} (m_0^b+ \delta m_b)\delta n_b$.    Here $\boldsymbol{m}_0$ is the average uniform
magnetization, which lies in the $y-z$ plane.  We assume
$m_0 \ll n_0$, so quantities are expressed to linear order
in $m_0$ whenever possible.
This gives
\begin{equation}
  \label{eq:142}
 \mathcal{H}'_{\rm
    tetra}(\mathbf{r})\approx\sum_{\alpha\beta}\mathcal{E}^{\alpha\beta}_{\mathbf{r}}
  \sum_{a,b=y,z}\sum_{\xi,\xi'=0,1}\lambda^{\alpha\beta}_{ab;\xi\xi'} n_0^{-\xi-\xi'}\eta_{a\xi{\mathbf{r}}}\eta_{b\xi'\mathbf{r}},
\end{equation}
where $\eta_{a0}=\delta n_a$ and $\eta_{a1}=\delta m_a$, 
and
\begin{eqnarray}
  \label{eq:15a}
\lambda_{ab;\xi\xi}^{\alpha\beta}&=&\Lambda_{ab}^{(\xi),\alpha\beta}-\delta_{ab}\Lambda_{xx}^{(0),\alpha\beta},\\
  \lambda_{ab;01}^{\alpha\beta}& = & \lambda_{ba;10}^{\alpha\beta}
                                     \nonumber \\
  &= &
                                        \frac{-1}{n_0}\left[
                                        m^a_0 \Lambda^{(1), \alpha\beta}_{bx} +
                                        \delta_{ab} m^{\overline a}_0
                                        \Lambda_{\overline a x}^{(1), \alpha\beta} + m^b_0
                                        \Lambda^{(0), \alpha\beta}_{ax}\right], \nonumber
\end{eqnarray}
where $\overline{y}=z$, $\overline{z}=y$ and we have associated
$\xi={\rm n}\Leftrightarrow \xi=0$ and
$\xi={\rm m}\Leftrightarrow \xi=1$ in $\Lambda^{(\xi)}$.

Here each $\Lambda^{(\xi)}$ tensor, which we define to be
symmetric in both $ab$ and $\alpha\beta$ variables, has seven independent coefficients, which we
call
$\Lambda_1^{(\xi)}=\Lambda_{xx}^{(\xi),xx}=\Lambda_{yy}^{(\xi),yy}$, 
  $\Lambda_2^{(\xi)}=\Lambda_{yy}^{(\xi),xx}=\Lambda_{xx}^{(\xi),yy}$,
  $\Lambda_3^{(\xi)}=\Lambda_{zz}^{(\xi),xx}=\Lambda_{zz}^{(\xi),yy}$, 
  $\Lambda_4^{(\xi)}=\Lambda_{xx}^{(\xi),zz}=\Lambda_{yy}^{(\xi),zz}$, 
  $\Lambda_5^{(\xi)}=\Lambda_{zz}^{(\xi),zz}$, 
  $\Lambda_6^{(\xi)}=\Lambda_{xy}^{(\xi),xy}=\Lambda_{xy}^{(\xi),yx}=\Lambda_{yx}^{(\xi),yx}=\Lambda_{yx}^{(\xi),xy}$, 
  $\Lambda_7^{(\xi)}=\Lambda_{xz}^{(\xi),xz}=\Lambda_{xz}^{(\xi),zx}=\Lambda_{zx}^{(\xi),zx}=\Lambda_{zx}^{(\xi),xz}
=\Lambda_{yz}^{(\xi),yz}=\Lambda_{yz}^{(\xi),zy}=\Lambda_{zy}^{(\xi),zy}=\Lambda_{zy}^{(\xi),yz}$. 
All other $\Lambda^{(\xi),\alpha\beta}_{ab}$ are zero. This is the most general coupling
allowed by the symmetries of the lattice and of the magnetic order.

To cast this in the form of Eq.~\eqref{eq:1} and Eq.~\eqref{eq:13}, we
insert the (very standard) free field expressions for the strain and magnetization fluctuations
in terms of phonon and magnon creation/annihilation operators,
respectively, into Eq.~\eqref{eq:142}.  For the strain,
\begin{align}
  \label{eq:33}
   & \mathcal{E}^{\mu\nu}(\mb{x})  = \\
  &\frac{1}{\sqrt{V}} \sum_{n\mb{k}}
  \frac{i/2}{\sqrt{2\rho_M \omega_{n\mb{k}}}}\left(a_{n\mb{k}}^{\vphantom\dagger} +
  a_{n,-\mb{k}}^\dagger\right) \left( k^\mu\varepsilon_{n\mb{k}}^\nu +k^\nu\varepsilon_{n\mb{k}}^\mu \right)e^{i\mb{k}\cdot\mb{x}},\nonumber
\end{align}
where $\rho_M$ is the mass density.  
For the magnetization densities,
diagonalization of the nonlinear sigma-model hamiltonian density
  \begin{eqnarray}
    \label{eq:34}
    \mathcal{H}_{m} &=& \frac{\rho}{2} \left ( |\ul{\bs \nabla} \delta
                        n_y|^2 + |\ul{\bs \nabla}
   \delta n_z|^2\right )\\
&+& \frac 1{2\chi}(\delta m_y^2+\delta m_z^2)+ \sum_{a=y,z}\frac
  {\chi\Delta_{a-1}^2}2 \delta n_{a}^2 \nonumber
  \end{eqnarray}
  yields
\begin{align}
  \label{eq:31}
    \eta_{a\xi\mathbf{r}}=\sum_\mathbf{p}\sum_{\ell=0,1}\sum_{q=\pm}U_{a\xi\ell q}(\mathbf{p})b^q_{\ell\mathbf{p}}e^{i\mathbf{p}\cdot\mathbf{r}},
\end{align}
with
\begin{align}
  \label{eq:32}
       U_{a\xi\ell q}(\mathbf{p})&=-\delta_{a-1,\ell-\overline{\xi}\;{\rm mod}2}F_{\xi q\ell}(\mathbf{p}),\\
        \label{eq:43b}
F_{\xi q\ell}(\mathbf{p})&=(iq)^{\overline{\xi}}(-1)^{\overline{\xi}\ell}(\chi\Omega_{\ell\mathbf{p}})^{\xi-\frac{1}{2}}.
\end{align}
We defined $\overline{\xi}=1-\xi$, i.e.\ $\overline{0}=1,\overline{1}=0$,
as well as $a=y\Leftrightarrow a-1=0$ and $a=z\Leftrightarrow a-1=1$.
In addition, in Eq.\,\eqref{eq:34}, $\rho$ is the antiferromagnetic spin stiffness, and $\chi$ is the magnetic susceptibility.
Inserting these definitions into Eq.~\eqref{eq:142}, some algebra
leads to the form of the text, with the coupling coefficients
\begin{widetext}
\begin{equation}
\label{eq:26}
   \mathcal{B}^{n,\ell_1\ell_2|q_1q_2q}_{\mathbf{k};\mathbf{p}} =
 \frac{iq}{2\sqrt{2M_{\rm uc}}} 
  \sum_{\xi\xi'} n_0^{-\xi-\xi'} \mathcal{L}_{n\mathbf{k};\xi,\xi'}^{q,\ell_1,\ell_2}\; F_{\xi q_1 \ell_1}\left(\mb{p}+\frac{q}{2}\mb{k}\right)
  F_{\xi' q_2\ell_2} \left(-\mb{p}+\frac{q}{2}\mb{k}\right),
\end{equation}
\end{widetext}
where
\begin{equation}
  \label{eq:27}
  \mathcal{L}_{n\mathbf{k};\xi,\xi'}^{q,\ell_1,\ell_2}=\sum_{\alpha,\beta=x,y,z}
  \hat{\lambda}^{\ell_1\ell_2;\alpha\beta}_{\xi\xi'}\;
  \frac{k^\alpha(\varepsilon_{n\mathbf{k}}^\beta)^q+k^\beta(\varepsilon_{n\mathbf{k}}^\alpha)^q}
  {\sqrt{\omega_{n\mathbf{k}}}},
\end{equation}
and $\hat{\lambda}_{\xi\xi'}^{\ell\ell';\alpha\beta}=\lambda^{\alpha\beta}_{\ell-\bar{\xi}\;{\rm
  mod}2,\ell'-\bar{\xi}'\;{\rm mod}2;\xi\xi'}$.

Note that the $\lambda_{mn}$ coefficients involved in the Hall conductivity,
  namely the $\bs \lambda_{ab;01}$ rank-2 tensors, are written explicitly in Eq.~\eqref{eq:15a}.
  They are proportional to the net magnetization $\bs m_0$, as is
  consistent with the fact that they are associated with a time-reversal breaking quantity.
One can also observe that they involve only the \emph{anisotropic} coefficients $\Lambda_{6,7}^{(\xi)}$,
which in a microscopic derivation arise from spin-orbit coupling, see Ref.~\cite{phononcompanion}.

\subsection{Numerical implementation}
\label{sec:numer-impl} 

In the numerical implementation, we use values of the parameters
roughly appropriate for CFTD, which we provide in Table~\ref{tab:table-params}.
The phonon polarization vectors $ \bs \varepsilon_{n,\mb k}$ are
chosen to form an orthonormal basis in which $\mb k$ points along the $[1,1,1]$
 axis, so that $\mb k \cdot \bs \varepsilon_{n,\mb k} = \frac{|\mb k|}{\sqrt 3} ~ \forall n$.

\begin{table}[htbp]
  \begin{tabular}{ccccccccccc}
\hline\hline
  $\,\frac{v_{\rm m}}{v_{\rm ph}}\,$ & $\chi \epsilon_0 \mathfrak{a}^2$ & $n_0$ &
                                                              $\frac{M_{\rm
                                                                          uc}v_{\rm
                                                                          ph}
                                                                          \mathfrak{a}}{\hbar}$
    & $m_0^x$
  &  $m_0^y$ & $m_0^z$ & $\frac{\Delta_0}{\epsilon_0}$ & $\frac{\Delta_1}{\epsilon_0}$ \\ \hline
  $\mt{2.5}$ & $\mt{0.19}$ & $\mt{1/2}$ &  $\mt{8\cdot 10^3}$ &  $\mt{0}$
  &  \begin{tabular}{@{}c@{}} $\mt{0.0}$ \\ $\mt{0.05}$\end{tabular} & \begin{tabular}{@{}c@{}} $\mt{0.05}$ \\ $\mt{0.0}$\end{tabular} & $\mt{0.2}$ & $\mt{0.04}$ \\
\hline\hline
  \end{tabular}
\begin{tabular}{c|ccccccc}
\hline\hline
  $\xi$ & $\Lambda^{(\xi)}_1$ &  $\Lambda^{(\xi)}_2$ & $\Lambda^{(\xi)}_3$ &
  $\Lambda^{(\xi)}_4$ & $\Lambda^{(\xi)}_5$ & $\Lambda^{(\xi)}_6$ & $\Lambda^{(\xi)}_7$\\ \hline
${\rm n} = 0$  & $\mt{12.0}$ & $\mt{10.0}$ & $\mt{14.0}$& $\mt{10.0}$ & $\mt{12.0}$ &$\mt{0.6}$ & $\mt{0.8}$ \\ \hline
 ${\rm m} = 1$ & $\mt{-10.0}$ & $\mt{-12.0}$& $\mt{-14.0}$ & $\mt{-12.0}$ & $\mt{-10.0}$ &$\mt{-0.8}$ & $\mt{-0.6}$\\ \hline\hline
\end{tabular}
\caption{Numerical values of the fixed dimensionless parameters used in all
  numerical evaluations.   The upper and lower entries for $m_0^y$ and
  $m_0^z$ correspond to the two cases for calculating $\varrho_H^{xy}$
  and $\varrho_H^{xz}$, respectively.  The couplings
  $\Lambda_i^{(\xi)}$ are given in units of
  $\epsilon_0/\mathfrak{a}$.  }
\label{tab:table-params}
\end{table}

\end{document}